\documentstyle[11pt]{article}
\def \AAP #1 #2 {{\em Astron. Astrophys.,\/} {\bf #1}, #2~}
\def \AAL #1 #2 {{\em Astron. Astrophys. Lett.,\/} {\bf #1}, L#2~}
\def \AAR #1 #2 {{\em Astron. Astrophys. Rev.,\/} {\bf #1}, #2~}
\def \AAS #1 #2 {{\em Astron. Astrophys. Suppl. Ser.,\/} {\bf #1}, #2~}
\def \AJ #1 #2 {{\em Astron. J.,\/} {\bf #1}, #2~}
\def \ANNREV #1 #2 {{\em Ann. Rev. Astron. Astrophys.,\/} {\bf #1}, #2~}
\def \APJ #1 #2 {{\em Astrophys. J.,\/} {\bf #1}, #2~}
\def \APJL #1 #2 {{\em Astrophys. J. Lett.,\/} {\bf #1}, L#2~}
\def \APJS #1 #2 {{\em Astrophys. J. Suppl.,\/} {\bf #1}, #2~}
\def \MN #1 #2 {{\em Mon. Not. R. Astr. Soc.,\/} {\bf #1}, #2~}
\def \PLR #1 #2 {{\em Phys. Lett. Rev.,\/} {\bf #1}, #2~}
\def \PASJ #1 #2 {{\em Publ. Astron. Soc. Japan,\/} {\bf #1}, #2~}
\def \PASP #1 #2 {{\em Publ. Astr. Soc. Pacific,\/} {\bf #1}, #2~}
\def \NAT #1 #2 {{\em Nature,\/} {\bf #1}, #2~}
\def \SCI #1 #2 {{\em Science,\/} {\bf #1}, #2~}
\def \FC #1 #2 {{\em Fund. Cosm. Phys.,\/} {\bf #1}, #2~}
\newcommand{\Ref}{\hangindent=20pt \hangafter=1 \noindent}
\newcommand{\StartRef}{\hyphenpenalty=10000 \raggedright
\parskip=0pt \parindent=0pt }

\def\gsim{\;\rlap{\lower 2.5pt
 \hbox{$\sim$}}\raise 1.5pt\hbox{$>$}\;}
\def\lsim{\;\rlap{\lower 2.5pt
   \hbox{$\sim$}}\raise 1.5pt\hbox{$<$}\;}
\def\ga{\;\rlap{\lower 2.5pt
 \hbox{$\sim$}}\raise 1.5pt\hbox{$>$}\;}
\def\la{\;\rlap{\lower 2.5pt
   \hbox{$\sim$}}\raise 1.5pt\hbox{$<$}\;}
\newcommand\beq{\begin{equation}}
\newcommand\eeq{\end{equation}}
\def\lya{Ly$\alpha$~}
\newcommand{\NarrowMargins}{
  \setlength{\oddsidemargin}{-0.2in}
  \setlength{\evensidemargin}{-0.2in}
  \setlength{\textwidth}{7.in}
  \setlength{\topmargin}{0.0in}
  \setlength{\textheight}{9.5in}
  \setlength{\headsep}{0.0in}
  \setlength{\parindent}{0.0in}
  \setlength{\parskip}{0.5cm}
  \setlength{\baselineskip}{0.5cm}
  \setlength{\headheight}{0.0in}}
\NarrowMargins
\input epsf.sty

\NarrowMargins
\begin{document}
\NarrowMargins
\addtolength{\topmargin}{1cm}

\vspace{4\baselineskip}
\noindent
{\LARGE\bf FORMATION OF THE FIRST STARS AND QUASARS}

\vspace{\baselineskip}
\noindent
Z. Haiman$^{1,2}$

\vspace{\baselineskip}
\noindent
$^{1}${\it Astronomy Department, Harvard University, 60 Garden Street, Cambridge, MA 02138, USA}

\vspace{-\baselineskip}
$^{2}${\it NASA/Fermilab Astrophysics Center, Fermi National Accelerator Laboratory, P.O. Box 500, Batavia, IL 60510, USA}

\vspace{2\baselineskip}
\noindent
ABSTRACT

\noindent
We examine various observable signatures of the first generation of
stars and low--luminosity quasars, including the metal enrichment,
radiation background, and dust opacity/emission that they produce.  We
calculate the formation history of collapsed baryonic halos, based on
an extension of the Press--Schechter formalism, incorporating the
effects of pressure and ${\rm H_2}$--dissociation.  We then use the
observed C/H ratio at $z$=3 in the Lyman--$\alpha$ forest clouds to
obtain an average the star formation efficiency in these halos.
Similarly, we fit the efficiency of black-hole formation, and the
shape of quasar light curves, to match the observed quasar luminosity
function (LF) between $z$=2--4, and use this fit to extrapolate the
quasar LF to faint magnitudes and high redshifts.  To be consistent
with the lack of faint point--sources in the Hubble Deep Field, we
impose a lower limit of $\sim75~{\rm km~s^{-1}}$ for the circular
velocities of halos harboring central black holes.

\noindent
We find that in a $\Lambda$CDM model, stars reionize the IGM at
$z_{\rm reion}$=9--13, and quasars at $z$=12.  Observationally,
$z_{\rm reion}$ can be measured by the forthcoming MAP and Planck
Surveyor satellites, via the damping of CMB anisotropies by $\sim$10\%
on small angular scales due to electron scattering.  We show that if
reionization occurs later, at $5\lsim z_{\rm reion}\lsim 10$, then it
can be measured from the spectra of individual sources. We also find
that the Next Generation Space Telescope will be able to directly
image about 1-40 star clusters, and a few faint quasars, from $z>10$
per square arcminute.  The amount of dust produced by the first
supernovae has an optical depth of $\tau$=0.1--1 towards high redshift
sources, and the reprocessed UV flux of stars and quasars distorts the
cosmic microwave background radiation (CMB) by a Compton $y$-parameter
comparable to the COBE limit, $y\sim1.5\times10^{-5}$.

\vspace{\baselineskip}
INTRODUCTION

One of the outstanding problems in cosmology is the nature of the
first generation of astrophysical objects which appeared when the
universe first transformed from its initial smooth state to its
current clumpy state.  Although we have observational data on bright
quasars and galaxies out to redshifts $z\sim5$ and on the linear
density fluctuations at redshift $z\sim10^3$, there is currently no
direct evidence as to when and how the first structures formed, and
what kind of objects were responsible for the end of the ``dark age''
of the universe (Rees~1996).

\addtolength{\topmargin}{-1cm}
\epsfysize=12cm 
\hspace{2.5cm}\epsfbox{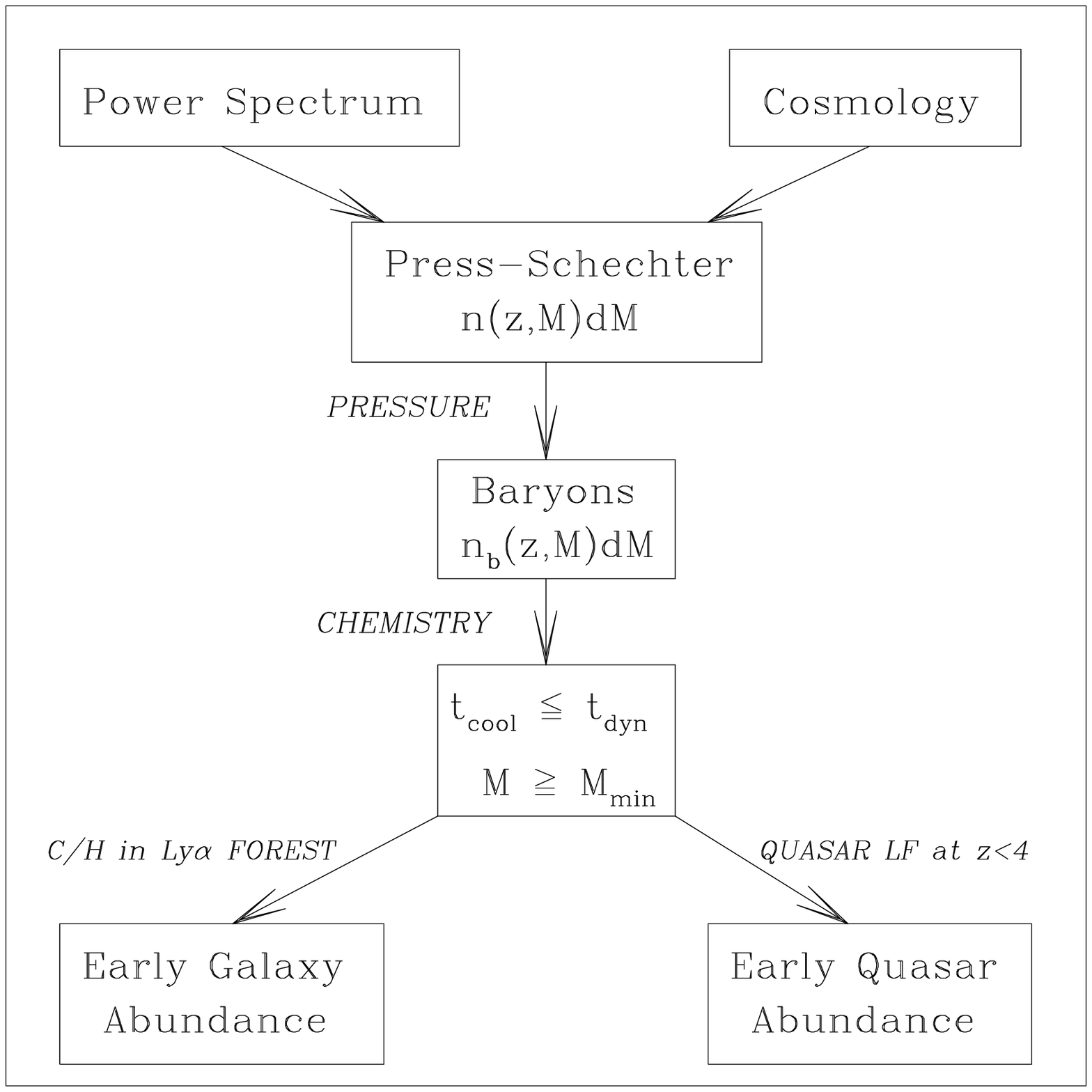} 
\begin{center}
Fig. 1: {\it Schematic View of Extrapolations to $z>5$.}
\end{center}

Popular Cold Dark Matter (CDM) models for structure formation predict
the appearance of the first baryonic objects with masses
$M\sim10^5{\rm M_\odot}$ at redshifts as high as $z\sim30$; objects
with progressively higher masses assemble later. Following
virialization, the gas in these objects can only continue to collapse
and fragment if it can cool on a timescale shorter than the Hubble
time.  In the metal--poor primordial gas, the only coolants that
satisfy this requirement are neutral atomic hydrogen (H) and molecular
hydrogen (${\rm H_2}$). However, ${\rm H_2}$ molecules are fragile,
and are easily photodissociated throughout the universe by trace
amounts of starlight (Haiman, Rees, \& Loeb 1997).  Hence, most of the
first stars are expected to form inside objects with virial
temperatures $T_{\rm vir}\gsim10^4$K.  Depending on the details of
their cooling and angular momentum transport, the gas in these objects
is expected to either fragment into stars, or form a central black
hole exhibiting quasar activity. Conversion of a small fraction ($\sim
1-10\%$) of the gas into stars or quasars could reionize the universe,
and strongly affect the entropy of the intergalactic medium.

Here, we explore the impact of early stars and quasars on the
reionization history of the universe, as well as their effects on the
CMB.  For both types of sources, we calibrate the total amount of
light they produce based on data from redshifts $z\la 5$. The
efficiency of early star formation is calibrated based on the observed
metallicity of the intergalactic medium (Tytler~et~al.~1995, Songaila
\& Cowie 1996), while the early quasars are constrained to match the
quasar luminosity function at redshifts $z\la 5$ (Pei 1995), as well
as data from the Hubble Deep Field (HDF) on faint point--sources.  We
focus on a particular cosmological model with a cosmological constant,
namely the ``concordance model'' of Ostriker \& Steinhardt (1995).
Within this model, we predict the number of high--redshift
star--clusters, and low-luminosity quasars, using the Press--Schechter
formalism (Press \& Schechter 1974).  For results in other
cosmological models, as well as a more detailed description of our
methods and results, we refer the reader to several papers (Haiman,
Rees \& Loeb 1997; Haiman \& Loeb 1997a,b; Loeb \& Haiman 1997; Haiman
\& Loeb 1998a,b; and Haiman, Madau \& Loeb 1998).

MODELING METHOD

The method described below is based on the Press--Schechter formalism,
and is depicted schematically in Figure~1.  The most important aspects
of our modeling are the inclusion of gas pressure, introducing the
requirement of efficient cooling, and calibrating the formation rate
of stars or black holes to the halo formation rate, using
observational data available at $z<5$.

\underline{Formation of Dark Halos}

The change in the comoving number density of dark halos with masses
between $M$ and $M+dM$, between redshifts $z$ and $z+dz$, is governed
by the derivative,
\beq
\frac{d^2 N(M,z)}{dMdz}= \frac{d}{dz}\frac{dN_{\rm ps}(M,z)}{dM}
\eeq 
where $dN_{\rm ps}/dM(M,z)$ is the Press-Schechter mass function.  The
actual halo formation rate is larger than the derivative ${d\over
dz}(dN_{\rm ps}/dM)$, since this derivative includes a negative
contribution from merging halos.  However, at high redshifts collapsed
objects are rare, and the merger probability is low. We have compared
the above expression with the more accurate result for the halo
formation rate given by Sasaki (1994), and confirmed that the
difference in the rates is negligible for the high redshifts and halo
masses under consideration here.  We therefore use Eq.~(1) to
describe the dark halo formation rate at high redshifts.

\underline{Effects of Gas Pressure}

Although we use the Press--Schechter mass--function to find the
abundance and mass distribution of dark matter halos, this does not
directly yield the corresponding mass--function of baryonic clouds.
Initially, most of the collapsed baryons are in low--mass systems near
the Jeans mass, and therefore the pressure of the baryons has a
significant effect on the overall collapsed fraction.  Effectively,
the collapse of the baryons within each halo is delayed relative to
the dark matter (Haiman, Thoul \& Loeb 1996), reducing the baryon
fraction of low--mass halos below the initial $\Omega_{\rm b}$.  We
obtained the exact collapse redshifts, and baryon fractions, of
spherically symmetric perturbations by following the motion of both
the baryonic and the dark matter shells with a one dimensional
hydrodynamics code (Haiman \& Loeb 1997a).  This produces a
one--to--one mapping of dark matter mass to baryon mass within each
halo, as a function of halo mass and collapse redshift, which we then
use to obtain $d^2 N_{\rm b}/dM_{\rm b}dz$, the rate of change of mass
function of gas clouds, as a function of cloud mass and redshift.

\underline{Effect of Cooling}

Following collapse and virialization, the gas within each cloud can
only continue to collapse if it can cool on a timescale shorter than
the Hubble time.  This is a necessary condition for a cloud to form
either stars or central black holes, and translates to a lower limit
on the cloud temperature, or the cloud mass.  In the metal--poor
primordial gas, there are two cooling agents that satisfy this
requirement: neutral atomic hydrogen (H) and molecular hydrogen (${\rm
H_2}$); the former is effective at $T_{\rm H}\ga 10^4$K, the latter at
$T_{\rm H_2}\ga 10^2$K.  Taking the temperature of each collapsed gas
cloud to be the virial temperature, these limits would imply the low
mass cutoffs of $M_{\rm H}\sim10^{8}{\rm M_{\odot}}[(1+z_{\rm
vir})/10]^{-3/2}$ or $M_{\rm H_2}\sim10^{5}{\rm M_{\odot}}[(1+z_{\rm
vir})/10]^{-3/2}$, respectively.  Since we found that ${\rm H_2}$
molecules are fragile, and are easily photodissociated throughout the
universe by trace amounts of starlight (Haiman, Rees, \& Loeb 1997),
we here assume that most of the first stars or quasars are form via
atomic H cooling, inside objects with masses $M\gsim M_{\rm H}$.

\underline{Star--Formation Efficiency}

One possible fate of each collapsed gas cloud that can cool and
continue to collapse is to fragment into stars.  We here ignore the
actual fragmentation process, and simply calibrate the fraction
$f_{\rm star}$ of the collapsed gas that must have been converted into
stars, by requiring it to reproduce the universal average C/H ratio,
inferred from CIV absorption lines in \lya~forest clouds.  We assume
that the carbon produced in the early star clusters is homogeneously
mixed with the rest of the baryons; incomplete mixing would
necessitate more stars than we derive.  The C/H ratio deduced from the
observations is between $10^{-3}$ and $10^{-2}$ of the solar value
(Songaila 1997).  We use tabulated $^{12}$C yields of stars with
various masses, and consider three different initial mass functions
(IMFs).  The uncertainty in the total carbon production is a factor of
$\sim$10; a factor of $\sim$3 is from the uncertainty in the carbon
yields of $3-8{\rm M_\odot}$ stars due to the unknown extent of hot
bottom burning (Renzini \& Voli 1981), and another factor of $\sim$3
is due to the difference between the Scalo and Miller--Scalo (1979)
IMFs.  To be conservative, we assume inefficient hot bottom burning,
i.e. maximum carbon yields.  Under these assumptions, we find that the
condition $10^{-3}{\rm [C/H]_\odot} < {\rm C/H} < 10^{-2}{\rm
[C/H]_\odot}$ translates into $1.7\%<f_{\rm star}<17\%$. Since the
collapsed fraction at $z=3$ is $\sim$50\%, the fraction of all baryons
in stars is $\sim$0.8--8\%.  A factor of $\sim$3 is included in this
number due to the finite time required to produce carbon inside the
stars; i.e. only a third of the total stellar carbon is produced and
ejected by $z=3$.

\underline{Black--Hole Formation Efficiency and Quasar Light--Curve}

Another possible fate of a gas cloud that can cool is to continue
collapsing, and to form a central black hole, exhibiting quasar--like
activity.  To quantify this scenario, here we assume that the
luminosity history of each black hole depends only on its mass, and
postulate the existence of a universal quasar light--curve in
Eddington units.  This approach is motivated by the fact that for a
sufficiently high fueling rate, quasars are likely to shine at their
maximum possible luminosity, which is some constant fraction of the
Eddington limit, for a time which is dictated by their final mass and
radiative efficiency.  We also assume that the final black hole mass
is a fixed fraction of the total halo mass, and allow this fraction to
be a free parameter.  Based on this minimal set of assumptions, we
demonstrated that there exists a universal light curve [$f(t)=(L_{\rm
Edd}/M_{\rm bh})\exp(-t/t_0)$, with $t_0=10^{5.82}$], for which the
Press--Schechter theory provides an excellent fit to the observed
evolution of the luminosity function of bright quasars between
redshifts $2.6<z<4.5$.  Furthermore, our fitting procedure results in
a black hole to halo mass ratio of $M_{\rm bh}/M_{\rm
gas}=10^{-3.2}\Omega_{\rm m}/\Omega_{\rm b}=5.4\times10^{-3}$, close
to the typical value $\sim6\times10^{-3}$ found in local galaxies
(Kormendy {\it {\it et al.}} 1997; Magorrian {\it et al.} 1998). Given
this ratio and the fitted light--curve, we then extrapolate the
observed LF to higher redshifts and low luminosities.  Note, however,
that our solution is not unique, and with the introduction of
additional free parameters, a non--linear (mass and redshift
dependent) black--hole mass to halo mass relation can also lead to
acceptable fits to the observed quasar LF (Haehnelt, Natarajan, \&
Rees 1998).  The actual formation process of low--luminosity quasars
was addressed in detail by Eisenstein \& Loeb (1995), and Loeb (1997).

\underline{Constraints from the Hubble Deep Field}

High resolution, deep imaging surveys can be used to set important
constraints on semi--analytical structure formation models. We have
found that the lack of unresolved B--band ``dropouts'' with $V>25$ mag
in the

\epsfysize=12cm 
\hspace{2.5cm}\epsfbox{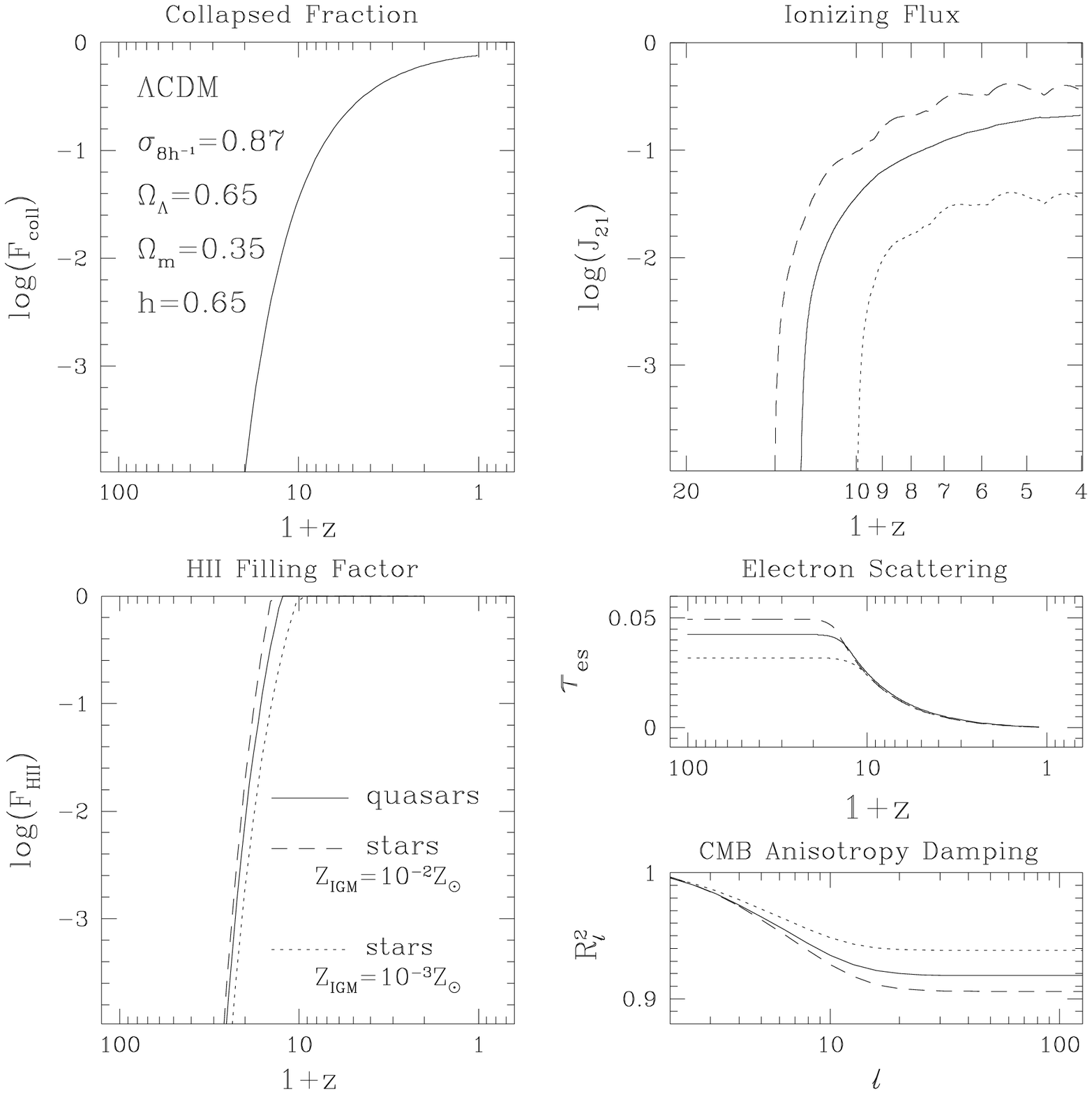} 

\noindent Fig. 2: {\it Reionization History.  Clockwise, the different panels
show: (i) the collapsed fraction of baryons; (ii) the background
flux at the Lyman edge; (iii) the volume filling
factor of ionized hydrogen; and (iv) the optical depth to electron
scattering, and the corresponding damping factor for the
power--spectrum decomposition of microwave anisotropies as a function
of the spherical harmonic index $l$.  The solid curves refer to
quasars, while the dotted/dashed curves correspond to stars
with low/high normalization for the star formation efficiency
(Haiman \& Loeb 1998a).}

\noindent Hubble Deep Field (HDF) appears to be inconsistent with the
expected number of quasars if massive black holes form with a constant
universal efficiency in all CDM halos, extending down to halos with
virial temperatures of $T_{\rm vir}=10^4$K as outlined above.  To
reconcile the models with the data, a mechanism is needed that
suppresses the formation of quasars in halos with circular velocities
$v_{\rm circ} \lsim 50-75~{\rm km~s^{-1}}$.  This feedback naturally
arises due to the photoionization heating of the gas by the UV
background.  We have considered several alternative effects that would
help reduce the quasar number counts, and find that these can not
alone account for the observed lack of detections (Haiman, Madau \&
Loeb 1998). We therefore impose a lower cutoff of $\sim75~{\rm
km~s^{-1}}$ for the circular velocities of halos harboring central
black holes.

RESULTS

\underline{Reionization}

Given the star--formation or quasar black--hole formation histories
obtained above, we derive the reionization history of the IGM by
following the radius of the expanding Str\"omgren sphere around each
source.  

\epsfysize=12cm 
\hspace{2.5cm}\epsfbox{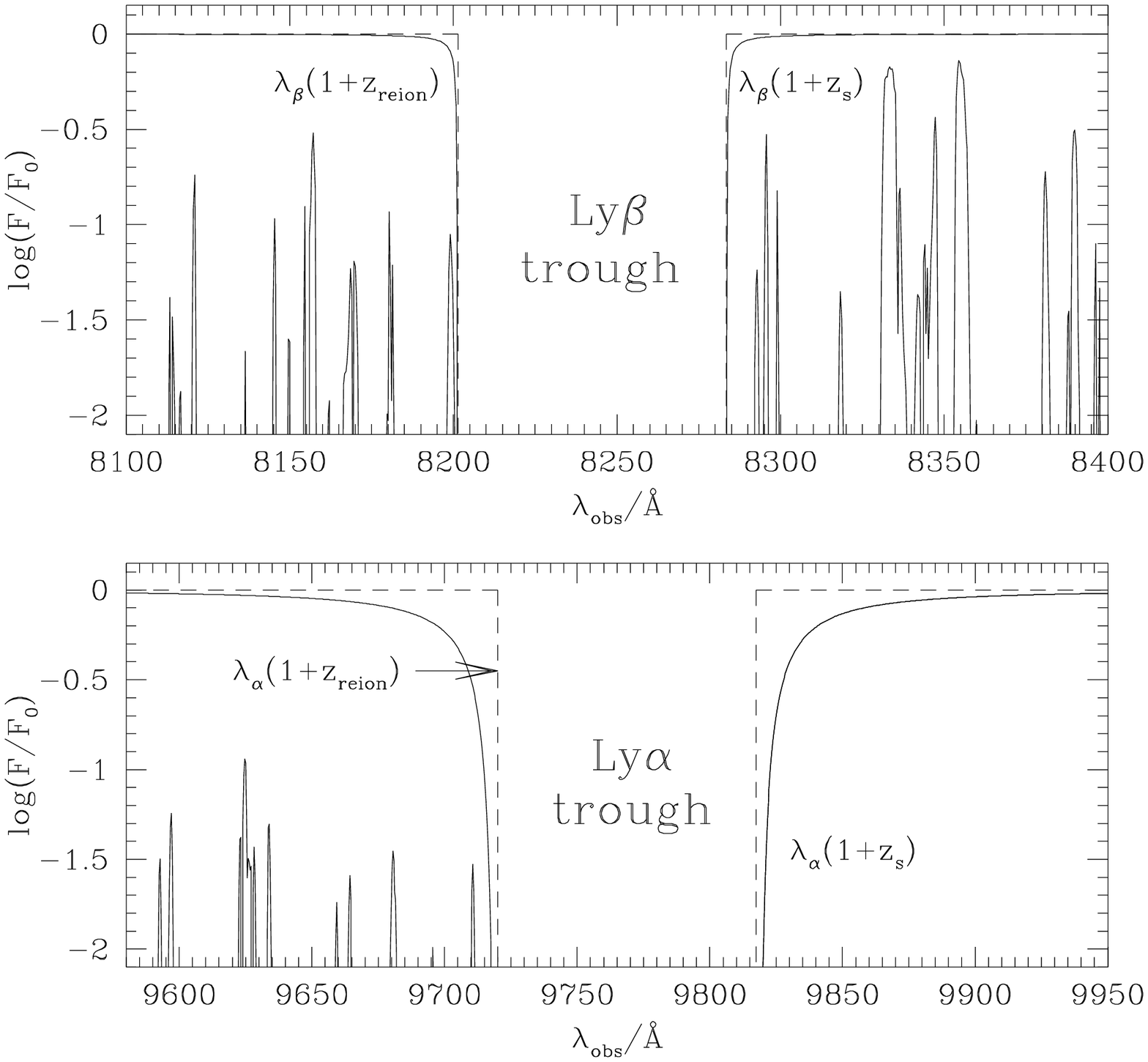} 

\noindent Fig. 3: {\it Spectrum of a source at $z_{\rm s}=7.08$, assuming
sudden reionization at a redshift $z_{\rm reion}=7$.  The solid curves
show the spectrum without absorption by the high--redshift \lya
forest, and the dashed lines show the spectrum when the damping wings
are also ignored.}

\noindent 
The reionization history depends on the time--dependent
production rate of ionizing photons, their escape fraction, 
and the
number of recombinations in the IGM, which are all functions of
redshift.  The ionizing photon rate per quasar follows from a median
quasar spectrum (Elvis~et~al.~1994) and the light--curve we derived.
The analogous rate per star follows from a time--dependent composite
stellar spectrum, constructed from standard stellar atmosphere atlases
(Kurucz 1993) and evolutionary tracks (Schaller~et~al~1992). We also
included the feedback from the fact that the collapse of new clouds is
suppressed in regions that are already ionized.

Figure~2 summarizes the resulting reionization histories of stars or
quasars (solid lines) in our $\Lambda$CDM cosmology. The results for
stars are shown in two cases, one with $Z_{\rm IGM}=10^{-2}{\rm
Z_\odot}$ (dashed lines) and the other with $Z_{\rm IGM}=10^{-3}{\rm
Z_\odot}$ (dotted lines), to bracket the allowed IGM metallicity
range.  The panels in Figure~2 show (clockwise) the collapsed fraction
of baryons; the evolution of the average flux, $J_{21}$ at the local
Lyman limit frequency, in units of $10^{-21}~{\rm
erg~s^{-1}~cm^{-2}~Hz^{-1}~sr^{-1}}$, the resulting evolution of the
ionized fraction of hydrogen, $F_{\rm HII}$, and the consequent
damping of the CMB anisotropies.  The dashed and dotted curves
indicate that stars ionize the IGM by a redshift $9\lsim z\lsim13$;
while the solid curve shows that quasars reionize the IGM at $z=11.5$.
This result can be understood in terms of the total number of
ionizing photons produced per unit halo mass: the
relative ratios of this number in the three cases are
$1\div0.37\div0.1$, respectively.  It is interesting to note, given
our quasar and stellar template spectra, that stars will not reionize
HeII, while quasars reionize HeII at slightly above the H reionization
redshift.

\epsfysize=11cm 
\hspace{3.cm}\epsfbox{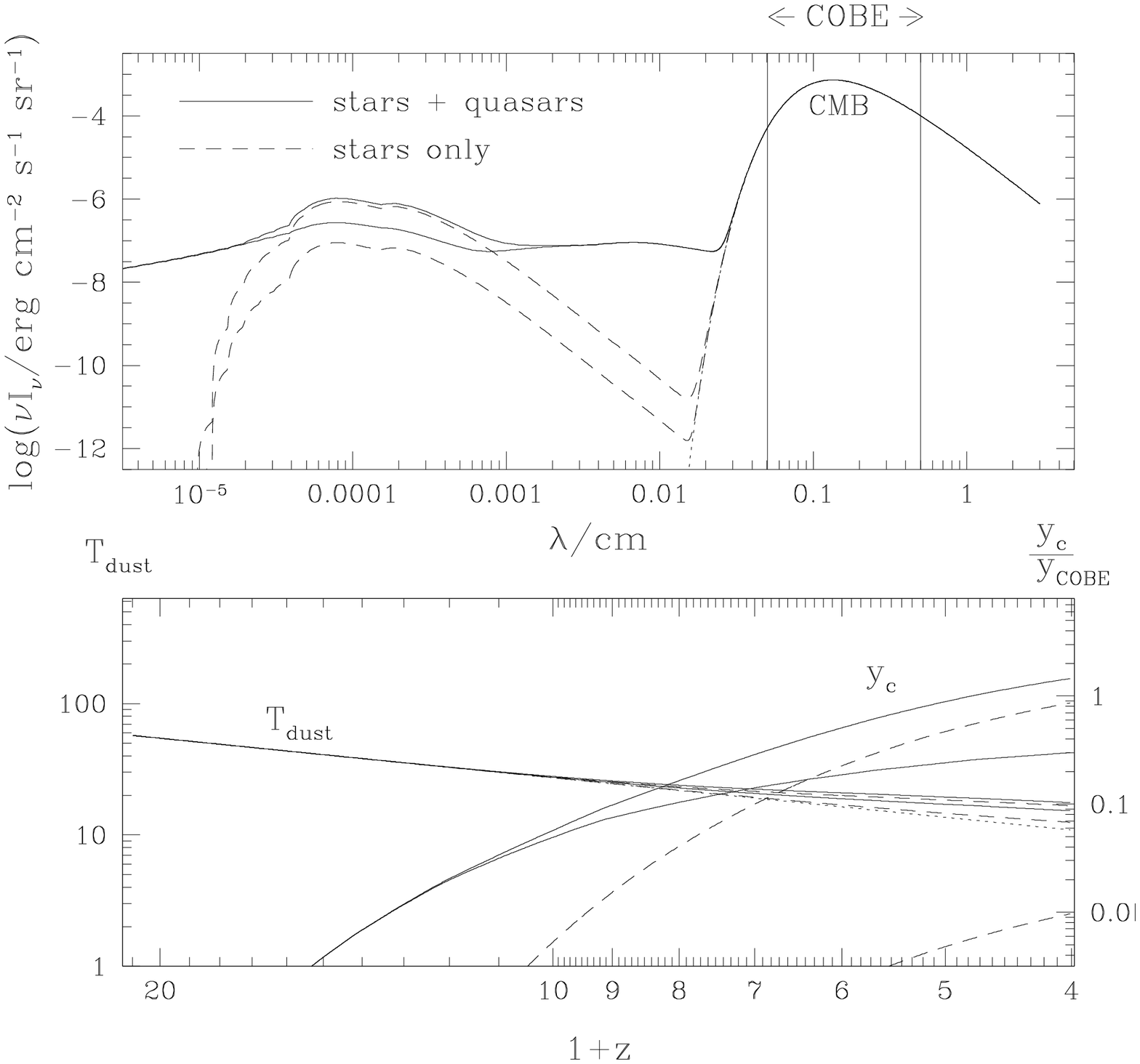} 
\begin{center}
Fig. 4: {\it Effect of Dust on the Background Flux.}
\end{center}

\noindent
In these scenarios the reionization redshift $z_{\rm reion}$ can be
inferred from the observations of the CMB (see next section). However
this method relies on post-reionization electron scattering, which has
a detectable optical depth only if $z_{\rm reion}\gsim 10$.  It is
therefore interesting to ask how $z_{\rm reion}$ can be measured in
general. In particular, for $z_{\rm reion}\lsim 10$ this would be
possible from the spectra of individual high--redshift sources.  The
spectrum of a source at a redshift $z_{\rm s}>z_{\rm reion}$ should
show a Gunn--Peterson (GP, 1965) trough due to absorption by the
neutral IGM at wavelengths shorter than the local \lya resonance at
the source, $\lambda_{\rm obs}<\lambda_\alpha(1+z_{\rm s})$. By
itself, the detection of such a trough would not uniquely establish
the fact that the source is located beyond $z_{\rm reion}$, since the
lack of any observed flux could be equally caused by (i) ionized
regions with some residual neutral fraction, (ii) individual damped
\lya absorbers, or (iii) line blanketing from lower column density
\lya forest absorbers.  On the other hand, for a source located at a
redshift $z_{\rm s}$ beyond but close to reionization, $(1+z_{\rm
reion}) < (1+z_{\rm s}) < \frac{32}{27} (1+z_{\rm reion})$, the GP
trough splits into disjoint Lyman $\alpha$, $\beta$, and possibly
higher Lyman series troughs, with some transmitted flux in between
these troughs. Although the transmitted flux is suppressed
considerably by the dense Ly$\alpha$ forest after reionization, it is
still detectable for sufficiently bright sources and can be used to
infer the reionization redshift.

As an example, we show in Figure~3 the simulated spectrum around the
Ly $\alpha$ and $\beta$ GP troughs of a source at redshift $z_{\rm
s}=7.08$, assuming reionization occurs suddenly at $z_{\rm reion}=7$.
We have included the effects of a mock catalog of Ly$\alpha$ absorbers
along the lines of sight, whose statistics were chosen to obey the
observational data at $z<4.3$ (Press \& Rybicki 1993).  Although the
continuum flux is strongly suppressed, the spectrum contains numerous
transmission features; these features are typically a few \AA~wide,
have a central intensity of a few percent of the underlying continuum,
and are separated by $\sim 10$\AA.  With a

\epsfysize=11cm 
\hspace{3cm}\epsfbox{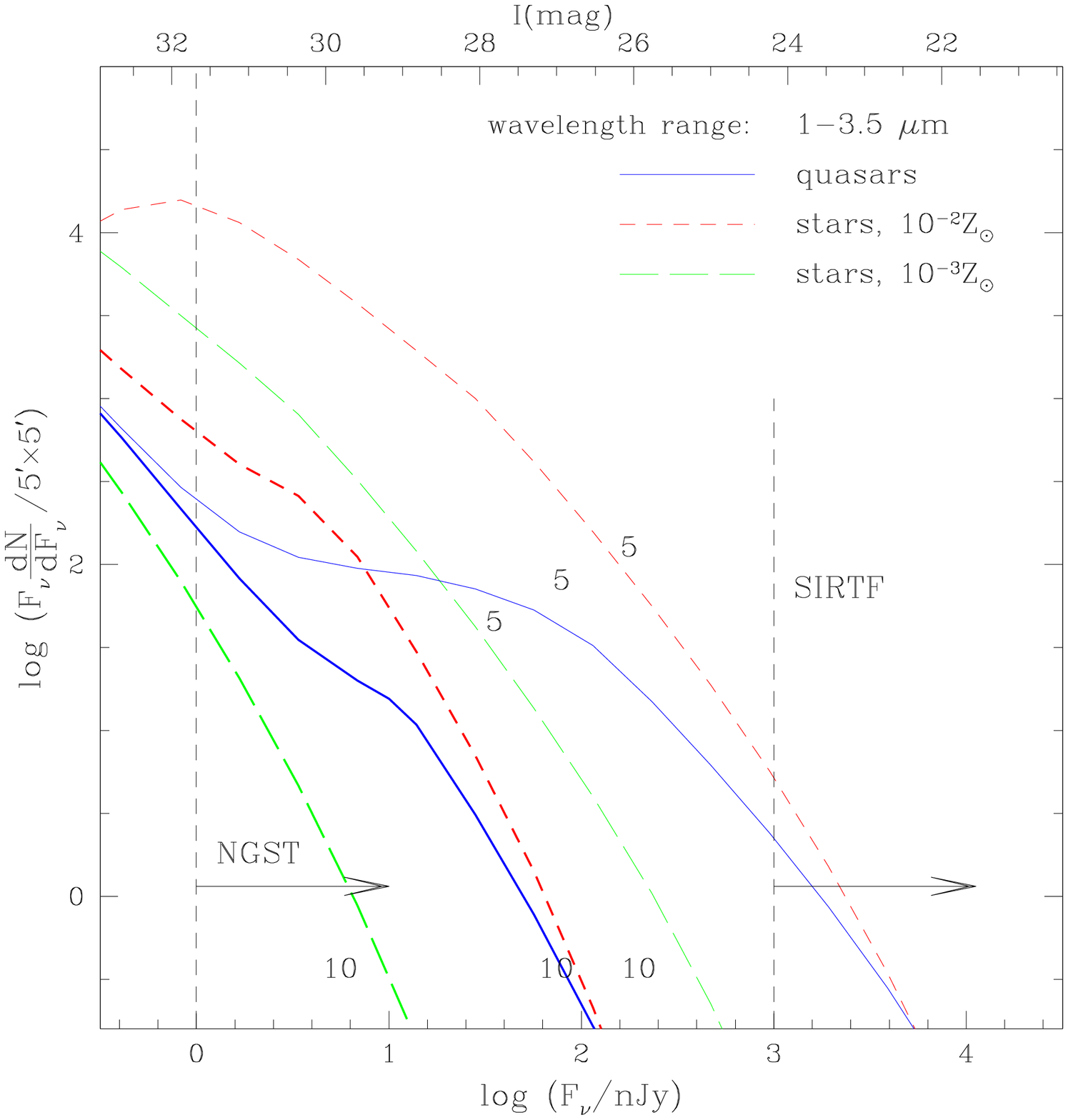} 

\noindent
Fig. 5: {\it Infrared Number Counts. The solid curves refer to
quasars, while the long(short) dashed curves correspond to star
clusters with low(high) normalization for the star formation efficiency.
The curves labeled ``5(10)'' show the cumulative number of objects
with redshifts above $z=5(10)$ (Haiman \& Loeb 1998a).}

\noindent sensitivity  reaching 1\% of
the continuum, the blue edges of the GP troughs could be identified to
a $\sim10$ \AA~accuracy, leading to a measurent of $z_{\rm reion}$ to
a $\sim10/8200\sim0.1\%$ precision.  Note that this accuracy is
similar to the level allowed by peculiar velocities of a few hundred
km/s, which could move either observed edge of the GP trough by $\sim
10$\AA.  Based on Figure~5 below, we expect that the Next Generation
Space Telescope would reach the spectroscopic sensitivity required for
the detection of sources suitable for this type of measurement (see
Haiman \& Loeb 1998b for details).

\underline{Signatures Imprinted on the Cosmic Microwave Background}

Reionization results in a reduction of the temperature anisotropies of
the CMB, due to Thomson scattering between the free electrons and CMB
photons.  Given the ionized fraction of hydrogen as a function of
redshift, the electron scattering optical depth ($\tau_{\rm es}$), as
well as the multiplicative anisotropy damping factor ($R^2_\ell$), as
a function of the spherical harmonic index $\ell$, can be readily
obtained (Hu \& White 1997). As shown in the lower right panel of
Figure~2, the amplitude of the anisotropies is reduced by an amount
between 6--10\% on small angular scales (large $\ell$). Although
small, this reduction is within the proposed sensitivities of the
future MAP and Planck satellite experiments, provided data on both
temperature and polarization anisotropies of the CMB is gathered (see
Table 2 in Zaldarriaga {\it et al.} 1997).

The stellar dust inevitably produced by the first type II supernovae
absorbs the UV emission from early stars and quasars, re--emits it at
longer wavelengths, and distorts the CMB spectrum.  We calculated this
spectral distortion assuming that each type II supernova yields ${\rm
0.3M_\odot}$ of dust with the wavelength-dependent opacity of Galactic
dust (Mathis 1990), which gets uniformly distributed within the
intergalactic medium.  The top panel of Figure~4 shows the resulting
total spectrum of the radiation background (CMB + direct stellar and
quasar emission + dust emission) at $z=3$ (more distortion could be
added between $0<z<3$ by dust and radiation from galaxies).  The
deviation from the pure 2.728$(1+z)$K blackbody shape is quantified by
the Compton $y$--parameter, whose redshift evolution is shown in the
bottom panel. Ignoring the UV flux from quasars, we obtain
$1.3\times10^{-7}<y_{\rm c}<1.2\times10^{-5}$ at $z=3$ (dashed lines),
just below the upper limit set by COBE, $y=1.5\times10^{-5}$
(Fixsen~et~al.~1996).  Adding the UV flux of quasars increases the
$y$--parameter to $4.1\times10^{-6}<y_{\rm c}<2.0\times10^{-5}$.  It
is interesting to note a substantial fraction ($\sim17$--$82$\%) of
this $y$--parameter results simply from the direct far-infrared flux
of early quasars, and should be present even in the absence of any
intergalactic dust.

\underline{Infrared Number Counts}

 Finally, we examine the feasibility of direct detection of the early
population of star clusters, and quasars, motivated by the Next
Generation Space Telescope ({\it NGST}).  {\it NGST} is scheduled for launch in
2007, and is expected to reach a sensitivity of $\sim 1$ nJy for
imaging in the wavelength range 1--3.5$\mu$m (Mather \& Stockman
1996). Figure~5 shows the predicted number counts, normalized to a
$5^{\prime}\times5^{\prime}$ field of view.  This figure shows
separately the number per logarithmic flux interval of all objects
with $z>5$ (thin lines), and with $z>10$ (thick lines).  The number of
detectable sources is high: {\it NGST} will be able to probe about $\sim100$
quasars at $z>10$, and $\sim200$ quasars at $z>5$ per field of view.
The bright--end tail of the number counts approximately follows the
power law $dN/dF_\nu\propto F_\nu^{-2.5}$.  The dashed lines show the
corresponding number counts of ``star--clusters'', i.e. assuming that
each halo shines due to a starburst that converts a fraction
0.017-0.17 of the gas into stars.  These indicate that {\it NGST} would
detect $\sim40-300$ star--clusters at $z>10$ per field of view, and
$\sim600-10^4$ clusters at $z>5$.  Unlike quasars, star clusters could
in principle be resolved, if they extend over a scale comparable to
the virial radius of the collapsed halo (Haiman \& Loeb 1997b).

CONCLUSIONS

Based on a combination of cosmological models with simple
phenomenological prescriptions for high--redshift star and quasar
black hole formation, we find that early stars and quasars would
reionize the IGM at high enough redshifts to be observable by the MAP
and Planck satellites via the corresponding reduction of the CMB
anisotropies. If the reionization redshift is lower, than it can be
inferred from the spectra of individual high--redshift sources.  The
Compton $y$--parameter for the spectral distortion of the CMB due to
early stellar dust is just below the existing COBE upper limit.  Most
directly, a typical image by {\it NGST} might reveal numerous early star
clusters and quasars.

ACKNOWLEDGEMENTS

I would like to thank my former advisor, Avi Loeb, for years of advice
and encouragement, and Anne Kinney for the opportunity to present my
work at the COSPAR assembly.

REFERENCES

\vspace{\baselineskip}
{
\StartRef

\Ref Eisenstein, D. J., and Loeb, A., \APJ 443 11 (1995).

\Ref Elvis, M., Wilkes, B. J., McDowell, J. C., Green, R. F., Bechtold, J.,  Willner, S. P., Oey, M. S., Polomski, E., and Cutri, R., \APJS 95 1 (1994).

\Ref Fixsen, D. J., Cheng, E. S., Gales, J. M., Mather, J. C., Shafer, R. A., and Wright, E. L., \APJ 473 576 (1996).

\Ref Gunn, J. E., \& Peterson, B. A., \APJ 142 1633 (1965).

\Ref Haehnelt, M. G., Natarajan, P., and Rees, M. J., preprint astro-ph/9712259 (1997).

\Ref Haehnelt, M. G., and Rees, M. J.: 1993, \MN 263 168 (1993).

\Ref Haiman, Z., and Loeb, A., \APJ 483 21 (1997a)

\Ref Haiman, Z., and Loeb, A., in  Proceedings of {\it Science with the Next Generation Space Telescope}, eds. E. Smith and A. Koratkar (1997b).

\Ref Haiman, Z., and Loeb, A., \APJ 503 505 (1998a).

\Ref Haiman, Z., and Loeb, A., {\it Astroph. J.}, submitted, preprint astro-ph/9807070 (1998b).

\Ref Haiman, Z., Madau, P., and Loeb, A., {\it Astroph. J.}, in press, preprint astro-ph/9805258 (1998).

\Ref Haiman, Z., Rees, M. J. R., and Loeb, A., \APJ 476 458 (1997).

\Ref Haiman, Z., Thoul, A., and Loeb, A., \APJ 464 523 (1996).

\Ref Hu, W., and White, M., \APJ 479 568 (1997).

\Ref Kormendy, J., Bender, R., Magorrian, J., Tremaine, S., Gebhardt, K., Richstone, D., Dressler, A., Faber, S. M., Grillmair, C., and Lauer, T. R., \APJL 482 139 (1997).

\Ref Kurucz, R., CD-ROM No. 13, ATLAS9 Stellar Atmosphere Programs (1993).

\Ref Loeb, A., in  Proceedings of {\it Science with the Next Generation Space Telescope}, eds. E. Smith and A. Koratkar (1997b).

\Ref Loeb, A., and Haiman, Z., \APJ 490 571 (1997).

\Ref Magorrian, J., {\it et al.}, \AA 115 2285 (1998).

\Ref Mather J \& Stockman, P., {\it STSci Newsletter} {\bf 13}, 15 (1996).

\Ref Mathis, J. S., \ANNREV 28 37 (1990).

\Ref Miller, G. E., and Scalo, J. M., \APJS 41 513 (1979).

\Ref Ostriker, J. P., and Steinhardt, P. J., \NAT 377 600 (1995).

\Ref Pei, Y. C., \APJ 438 623 (1995).

\Ref Press, W. H., and Rybicki, G. B., \APJ 418 585 (1993).

\Ref Press, W. H., and Schechter, P. L., \APJ 181 425 (1974).

\Ref Rees, M. J., preprint astro-ph/9608196 (1996).

\Ref Renzini, A., and Voli, M., \AA 94 175 (1981).

\Ref Sasaki, S., \PASJ 46 427 (1994).

\Ref Scalo, J. M., \FC 11 1 (1986).

\Ref Schaller, G., Schaerer, D., Meynet, G., and Maeder, A., \AAS 96 269 (1992).

\Ref Songaila, A., \APJL 490 1 (1997).

\Ref Songaila, A., and Cowie, L. L., \AJ 112 335 (1996).

\Ref Tytler, D. {\it et al.}, in {\it QSO Absorption Lines}, ed. G. Meylan ed., Springer, p.289 (1995).

\Ref Zaldarriaga, M., Spergel, D., and Seljak, U., \APJ 488 1 (1997).

}

\end{document}